\documentclass[conference]{IEEEtran}  

\usepackage{color}
\usepackage{soul}
\usepackage{siunitx}
\usepackage{amsmath}
\usepackage{amssymb}
\usepackage{booktabs}
\usepackage[caption=false]{subfig}
\usepackage[pdftex]{graphicx}
\usepackage{url}
\usepackage{tikz}
\usepackage{textcomp}
\usepackage{hyperref}
\usepackage{lipsum}

\newcommand\copyrighttext{%
  \footnotesize \textcopyright 2024 IEEE.  Personal use of this material is permitted.  Permission from IEEE must be obtained for all other uses, in any current or future media, including reprinting/republishing this material for advertising or promotional purposes, creating new collective works, for resale or redistribution to servers or lists, or reuse of any copyrighted component of this work in other works.
  DOI: \href{https://doi.org/10.1109/CAMAD62243.2024.10942683}{10.1109/CAMAD62243.2024.10942683}}
\newcommand\copyrightnotice{%
\begin{tikzpicture}[remember picture,overlay]
\node[anchor=south,yshift=10pt] at (current page.south) {\fbox{\parbox{\dimexpr\textwidth-\fboxsep-\fboxrule\relax}{\copyrighttext}}};
\end{tikzpicture}%
}

\usepackage[left=1.01in,right=1.01in,top=0.75in,bottom=1.05in]{geometry}

\title{A Simulator for FANETs Using 5G Vehicle-to-Everything Communications and Named-Data Networking}

\author{%
  \IEEEauthorblockN{José Manuel Rúa-Estévez\IEEEauthorrefmark{1}, Alicia Meleiro-Estévez\IEEEauthorrefmark{1}, Pablo Fondo-Ferreiro\IEEEauthorrefmark{1}, Felipe Gil-Castiñeira\IEEEauthorrefmark{1},\\Brais Sánchez-Rama\IEEEauthorrefmark{2}, Lois Gomez-Gonzalez\IEEEauthorrefmark{2}}
  \vspace{0.02in}
  \IEEEauthorblockA{\IEEEauthorrefmark{1}atlanTTic, University of Vigo\\36310 Vigo, Spain}
  \vspace{0.01in}
  \IEEEauthorblockA{\IEEEauthorrefmark{2}CENTUM Research \& Technology,\\ 36310 Vigo, Spain}
  \vspace{0.01in}
  E-mails:~\url{{jmrua, ameleiro, pfondo, xil}@gti.uvigo.es},~\url{{brais.sanchez, lois.gomez}}@centum-rt.com
}

\begin{document}
\maketitle

\copyrightnotice

\begin{abstract}
This work presents a simulator designed for the validation, evaluation, and demonstration of flying ad-hoc networks (FANETs) using 5G vehicle-to-everything (V2X) communications and the named-data networking (NDN) paradigm. The simulator integrates the ns-3 network simulator and the Zenoh NDN protocol, enabling realistic testing of applications that involve the multi-hop communication among multiple unmanned aerial vehicles (UAVs).
\end{abstract}


\section{Introduction}
\label{sec:introduction}

Flying ad-hoc networks (FANETs) consist of multiple unmanned aerial vehicles (UAVs) that communicate with each other and with ground control stations (GCSs) to execute tasks cooperatively. In recent years, the number of use cases involving FANETs has grown significantly, encompassing distributed applications such as surveillance, delivery services, environmental monitoring, or disaster management~\cite{zhou2020uav}. Communication in these highly dynamic environments poses significant challenges, mainly motivated by the high mobility of the drones, which leads to continuous changes in the network topology~\cite{chriki2019fanet}.

Cellular vehicle-to-everything (C-V2X) technology is seen as a promising solution for addressing the challenges in FANETs, as it enables high data rates, ultra-low latency, and scalability while ensuring secure communications~\cite{mir2020enabling}. 5G sidelink enables direct communication between devices without relying on infrastructure, which is specially suited for dynamic network topologies~\cite{lien20203gpp}.

In this work, we present a simulator for the validation, evaluation, and demonstration of 5G V2X communication standards in the context of FANETs. This simulator enables testing the communications of a set of UAVs using the 5G sidelink channel to establish mesh networks with multi-hop capabilities. It is used to analyze the capacity of named data networking (NDN)~\cite{zhang2014named} to support communications in FANETs. NDN is a data-centric communication paradigm that is based on the publication and retrieval of data instead of communication between IP addresses. This paradigm adapts well to FANETs since it allows UAVs to communicate without needing to know their IP addresses, better accommodating the dynamic nature of topologies in FANETs.

This simulator is composed of a series of virtual machines (VMs) that each represent a node in the network (e.g., UAV or GCS). Each VM implements the NDN and application stacks of a drone, generating network traffic that traverses the simulated network.

The traffic generated by the VMs is routed through a simulated wireless network that connects them, simulates the mobility of each node, and reproduces realistic 5G V2X communication channels.

\section{Problem statement}
\label{sec:problem-statement}

We consider a FANET composed of dynamic nodes (e.g., UAVs) and static nodes (such as GCSs). These nodes execute a cooperative application that requires inter-node communication. Each device is equipped with 5G sidelink communication capabilities, facilitating direct wireless communication among them. To extend coverage, we consider that multi-hop communications may take place through the FANET,  using devices within range as relays for end-to-end communications. The communications  following a publish-subscribe paradigm using NDN.

Our goal is to provide a simulator capable of validating the operation of FANET applications and analyzing communication performance under realistic mobility and network conditions.

\section{Proposed solution}
\label{sec:proposed-solution}

We propose a network simulator capable of emulating FANET applications while simulating custom mobility behaviors and realistic 5G sidelink communications.

\begin{figure}[htb!]
\centering
\includegraphics[width=0.99\columnwidth]{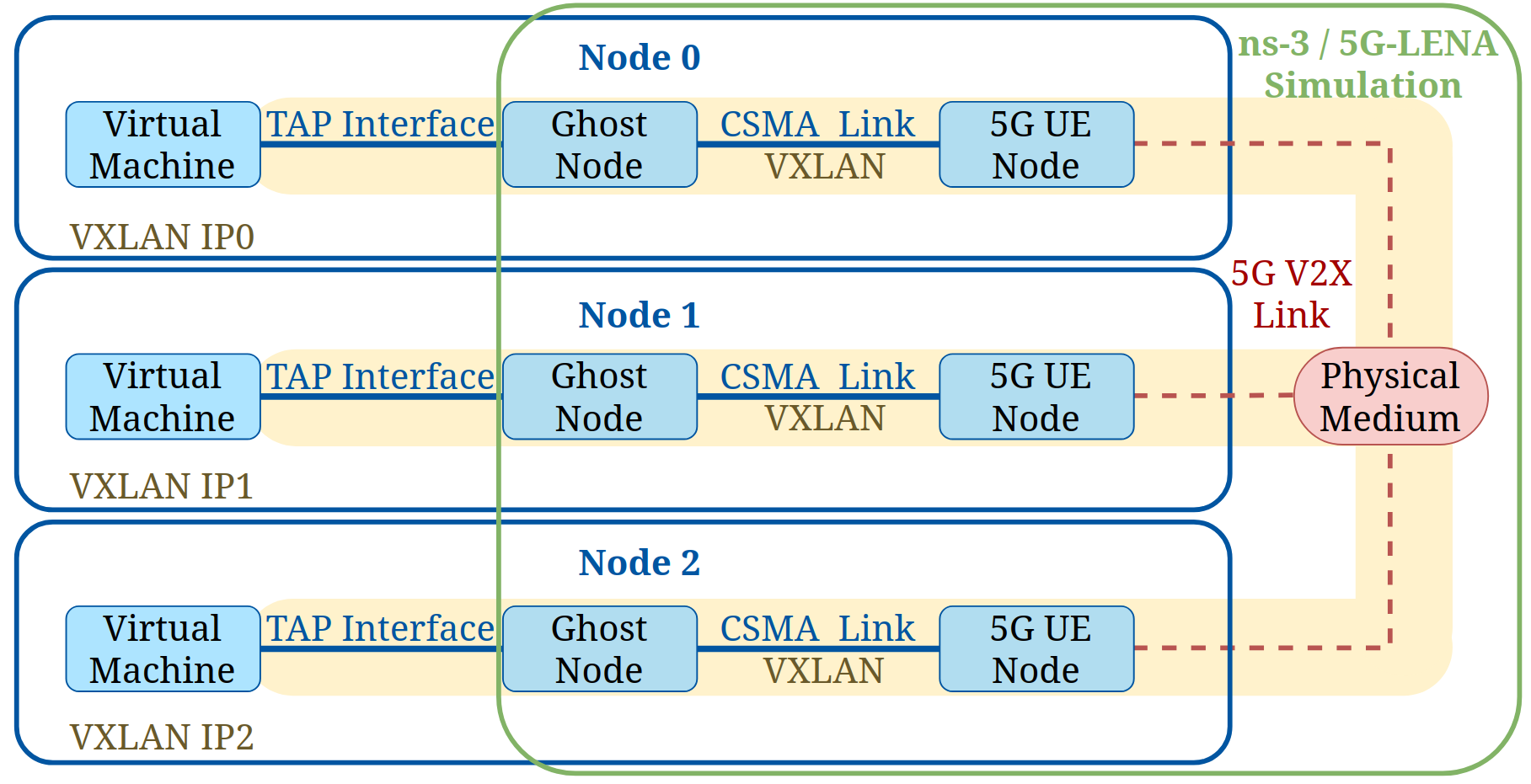}
\caption{Simulator architecture}
\label{fig:arquitectura}
\end{figure}

The simulator consists of a series of nodes interconnected via point-to-point links within  the ns-3  discrete-event network simulator~\cite{ns-3}. The 5G-LENA module of New Radio (NR) V2X is used in ns-3 for the simulation of sidelink communications between devices. 

The network simulator has limitations in implementing the application and NDN stacks. Therefore, we integrated these stacks into containers that execute code equivalent to what runs on the UAV on-board computer or the GCS computer. These containers are integrated with the emulated network through TUN/TAP network interfaces, as depicted in Fig.~\ref{fig:arquitectura}.

After entering the simulated network, traffic is intercepted by an intermediate ns-3 Carrier Sense Multiple Access (CSMA) node. This node acts as a transparent bridge between the host interface and the simulated node due to implementation limitations preventing direct connection of certain node types to the TUN/TAP interface. The ns-3 CSMA node ensures traffic passes through without introducing any additional effects.

The NDN communication paradigm is implemented using Zenoh~\cite{zenoh}, a protocol enabling efficient and scalable data publication and subscription~\cite{lopez2023decentralized,lopez2024unleashing}. Zenoh facilitates node publication and subscription in namespaces without requiring brokers, unlike protocols like Message Queuing Telemetry Transport (MQTT). It supports decentralized discovery of nodes and message routing.

Despite their transparency, intermediate nodes in the simulator introduce additional hops that do not exist in the real environment. Zenoh's Peer to Peer mode, which uses multicast traffic with a time-to-live (TTL) of 1 for local peer discovery, conflicts with this setup because VMs are located in different subnets. To address this challenge, a virtual extensible local area network (VXLAN) overlay network was implemented. Each VM creates a virtual interface within a shared IP address range, enabling successful Zenoh session discovery and association.

\section{Results}
\label{sec:results}

The developed simulator enables the validation of the operation of FANET applications in realistic mobility scenarios while using 5G sidelink communications.

We used the simulator to measure different metrics related to network performance, such as throughput, network latency, and jitter, and some metrics related to the Zenoh protocol. The simulator can also be used to evaluate the performance of different routing protocols for FANET applications.

\section{Conclusions}
\label{sec:conclusions}

In this work, we presented a new simulator for the validation and evaluation of a flexible and scalable platform for conducting exhaustive tests of communication protocols, routing algorithms, and other relevant technologies for FANETs based on 5G V2X. This is fundamental to ensure the successful deployment and reliable operation of these networks in real-world applications.

The simulator allowed us to validate the use of the Zenoh protocol for enabling multi-hop communications, increasing the robustness of the communication network. The demonstrator can be showcased in a single laptop and no special requirements are needed.

\section*{Acknowledgements}

This work was supported in part by Xunta de Galicia (Spain) under grants ED481B-2022-019 and ED431C 2022/04; and  Ministerio de Ciencia e Innovación (Spain) grant PID2020-116329GB-C21.

\bibliographystyle{IEEEtran}
\bibliography{IEEEfull, main.bib}

\end{document}